**Comment on "Nonlocal Effects on the Magnetic Penetration Depth in $d-$ wave Superconductors"**

It is generally believed that $d-$pairing superconductors display a linear temperature dependence of the magnetic penetration depth (MPD), $\lambda(T) - \lambda(0) \sim T$, at low temperatures. There are, however, experimental indications of a $T^2$-dependence of MPD in high$-T_c$ superconductors at temperatures below some temperature $T^*$. In a recent Letter [1] Kosztin and Leggett explained this behavior in terms of nonlocal effects in the electrodynamics. They showed that in clean superconductors with $d-$pairing surface effects lead to a $T^2$-dependence of $\lambda_{ab}(T)$, as extracted from optical and microwave experiments with the magnetic field orientated along the $c-$ direction, while the temperature dependence of $\lambda_c(T)$ remains linear. There exist, however, other measurement techniques of MPD (direct static magnetic measurements, the lower critical magnetic field $B_{c1}$, vortex properties, muon spin relaxation, etc.) to which the arguments put forward in [1] cannot be applied. In the following we show that $\lambda(T) - \lambda(0) \sim T^n$ with $n > 1$ is a consequence of a *general thermodynamics principle.*

For simplicity let us consider a uniform system where all properties depend on coordinates $\mathbf{r} - \mathbf{r}'$ only. The current-current correlator,

$$\eta(\mathbf{k},\omega) = k^2 - \frac{\omega^2}{c^2}\varepsilon_{tr}(\mathbf{k},\omega)$$

connects the vector potential $\mathbf{A}(\mathbf{k},\omega)$ to the external current $\mathbf{j}_{ext}(\mathbf{k},\omega)$ via

$$\eta(\mathbf{k},\omega)\mathbf{A}(\mathbf{k},\omega) = \frac{4\pi}{c}\mathbf{j}_{ext}(\mathbf{k},\omega) \quad (1)$$

In turn, the transversal dielectric function, $\varepsilon_{tr}(\mathbf{k},\omega)$, is related to the electromagnetic kernel $Q(\mathbf{k},\omega)$ by the relation $\varepsilon_{tr}(\mathbf{k},\omega) = 1 - \frac{4\pi Q(\mathbf{k},\omega)}{\omega^2}$. In the static case the additional free energy due to the presence of an external current $\mathbf{j}_{ext}(\mathbf{k})$ (we use a transversal gauge) can be written in the following form [2]:

$$\mathcal{F} = \frac{1}{8\pi}\int \frac{d^3k}{(2\pi)^3}\, \eta(\mathbf{k},\omega=0)\, |\mathbf{A}(\mathbf{k},\omega=0)|^2 \quad (2)$$

The definition of the operator of inverse MPD is

$$\frac{1}{\lambda^2(\mathbf{k},T)} = \lim_{\omega \to 0}\frac{\omega^2}{c^2}\{1 - \operatorname{Re}\varepsilon_{tr}(\mathbf{k},\omega)\} \equiv \frac{4\pi}{c^2}Q(\mathbf{k},\omega=0) \quad (3)$$

By using these relations and Maxwell's equations the free energy $\mathcal{F}$ can be rewritten in the form:

$$\mathcal{F} = \frac{1}{8\pi}\int \frac{d^3k}{(2\pi)^3}\left[k^2 + \frac{1}{\lambda^2(\mathbf{k},T)}\right]\frac{|\mathbf{k}\times\mathbf{B}(\mathbf{k};T)|^2}{k^4} \quad (4)$$

Here $\mathbf{B}(\mathbf{k},T)$ is the (temperature dependent) induced magnetic field and satisfies the equation:

$$(k^2 + \frac{1}{\lambda^2(\mathbf{k},T)})\mathbf{B}(\mathbf{k};T) = \frac{4\pi}{c}[i\,\mathbf{k}\times\mathbf{j}_{ext}(\mathbf{k})] \quad (5)$$

Differentiating Eq.(4) with respect to temperature we get an expression for the entropy:

$$S(T) = -\frac{\partial \mathcal{F}}{\partial T} = \frac{1}{8\pi}\int\frac{d^3k}{(2\pi)^3}\frac{\partial}{\partial T}\left[\frac{1}{\lambda^2(\mathbf{k},T)}\right]\frac{|\mathbf{B}(\mathbf{k};T)|^2}{k^2} \quad (6)$$

According to the Nernst principle (third law) the entropy should vanish in the limit $T \to 0$. From the positivity of the integrand we must conclude

$$\lim_{T\to 0}\frac{\partial \lambda(\mathbf{k},T)}{\partial T} = 0 \quad (7)$$

The argument can be extended to any nonuniform system.

We see that the vanishing of the first derivative of MPD for $T \to 0$ is a consequence of a general principle of thermodynamics (third law). The value of $T^*$ below which a $T^2$-dependence may be observed depends on the exact physical mechanism. It may be nonlocality, it may be the effect of impurities (as proposed in Ref. [3]). It may be also a consequence of collective excitations.

N. Schopohl[a] and O.V. Dolgov[a,b]
  a) Institut für Theoretische Physik
  Auf der Morgenstelle 14
  D-72076 Tübingen, Germany
  b) P.N. Lebedev Physical Institute
  Moscow, Russia

November 4, 1997

1